\documentclass[]{spie}  

\usepackage{amsmath,amsfonts,amssymb}
\usepackage{graphicx}
\usepackage[colorlinks=true, allcolors=blue]{hyperref}
\usepackage{booktabs}
\usepackage[final]{microtype}
\usepackage{mathtools}

\PassOptionsToPackage{compress, sort}{cite}

\title{Bi-Directional MS Lesion Filling and Synthesis Using Denoising Diffusion Implicit Model-based Lesion Repainting}

\author[a]{Jinwei~Zhang$^*$}
\author[b]{Lianrui~Zuo$^*$}
\author[b]{Yihao~Liu}
\author[c]{Samuel~Remedios}
\author[b]{\\Bennett~A.~Landman}
\author[a]{Jerry~L.~Prince}
\author[a]{Aaron~Carass}
\affil[a]{Image Analysis and Communications Laboratory, Department~of~Electrical~and~Computer~Engineering, Johns~Hopkins~University,~Baltimore,~MD~21218,~United~States\vspace*{0.4em}}
\affil[b]{Department of Electrical and Computer Engineering, Vanderbilt~University,~Nashville,~TN~37235,~United~States\vspace*{0.4em}}
\affil[c]{Department of Computer Science, Johns Hopkins University, Baltimore,~MD~21218,~United~States}

\authorinfo{* Co-first author with equal contribution. \\Further author information: Send correspondence to Jinwei~Zhang. E-mail: \url{jwzhang@jhu.edu}}

\begin{document} 
\maketitle

\begin{abstract}
Automatic magnetic resonance~(MR) image processing pipelines are widely used to study people with multiple sclerosis~(PwMS), encompassing tasks such as lesion segmentation and brain parcellation. 
However, the presence of lesion often complicates these analysis, particularly in brain parcellation. 
Lesion filling is commonly used to mitigate this issue, but existing lesion filling algorithms often fall short in accurately reconstructing realistic lesion-free images, which are vital for consistent downstream analysis. 
Additionally, the performance of lesion segmentation algorithms is often limited by insufficient data with lesion delineation as training labels.
In this paper, we propose a novel approach leveraging Denoising Diffusion Implicit Models~(DDIMs) for both MS lesion filling and synthesis based on image inpainting. 
Our modified DDIM architecture, once trained, enables both MS lesion filing and synthesis. 
Specifically, it can generate lesion-free T1-weighted or FLAIR images from those containing lesions; Or it can add lesions to T1-weighted or FLAIR images of healthy subjects. 
The former is essential for downstream analyses that require lesion-free images, while the latter is valuable for augmenting training datasets for lesion segmentation tasks. 
We validate our approach through initial experiments in this paper and demonstrate promising results in both lesion filling and synthesis, paving the way for future work.
Our code is available at \url{https://gitlab.com/IACL/UponAcceptance}.
\end{abstract}


\section{INTRODUCTION}
\label{s:intro}
Multiple sclerosis (MS) is a chronic central nervous system disease characterized by inflammation, demyelination, and axonal damage~\cite{haider2016topograpy}.
Magnetic resonance (MR) imaging is a common imaging modality used to diagnose and monitor MS lesions.
Automatic MR image processing pipelines facilitate the study of people with multiple sclerosis (PwMS) at scale.
However, limitations exist at several steps within these pipelines related to MS lesions, such as insufficient training data with manual lesion mask delineation for developing reliable automatic lesion segmentation models, and difficulties in generating realistic lesion-free images during the lesion-filling step, which are crucial for accurate downstream brain parcellation.

Denoising diffusion probabilistic models (DDPMs) are state-of-the-art deep generative models that synthesize realistic data by converting random noise into structured output through many iterations of a reverse Markov process~\cite{ho2020denoising}.
In this work, we explore the feasibility of applying DDPMs to address MS lesion filling and synthesis in multi-contrast magnetic resonance (MR) images.
Specifically, we propose a bi-directional MS lesion filling and synthesis method using T1-weighted~(T1w) and fluid attenuated inversion recovery~(FLAIR) MR images, inspired by a natural image inpainting approach using a DDPM~\cite{lugmayr2022repaint}.
Our approach includes DDIM-based inference acceleration with a few modifications, taking into account the non-Markovian nature of DDIMs~\cite{song2021iclr}.
We validate the performance of the proposed lesion filling and synthesis in downstream whole-brain parcellation and lesion segmentation tasks.

\section{METHOD}
\label{s:method}
\begin{figure}
    \centering
    \includegraphics[width=0.8\linewidth]{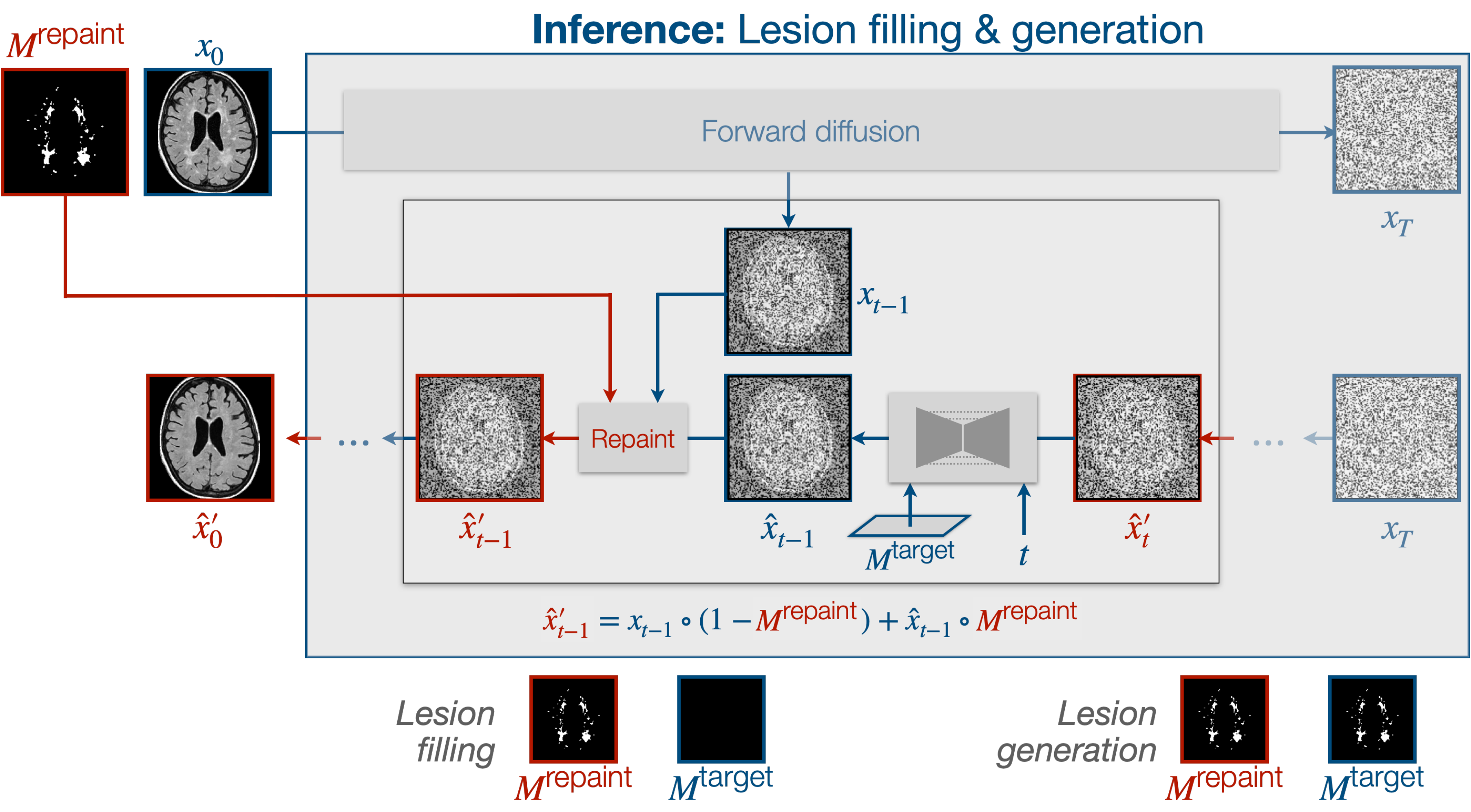}
    \caption{During inference, the DDPM is guided by the target lesion mask $M^\text{target}$ either generating lesion-free or synthetic lesion images. The mask $M^\text{repaint}$ controls the pixels to replace and preserve during the reverse process.}
    \label{fig:figure1}
\end{figure}
MS lesion filling and generation can both be formalized as image inpainting problems, where the objective is to modify pixel characteristics within a given region of interest.
The two primary tasks of the DDPM in this context are: 1)~to discern when to perform lesion filling versus lesion generation, and 2)~to replace lesion voxels with healthy-looking voxels for lesion filling or vice-versa for lesion generation. 

\subsection{Training DDPM with lesion mask guidance}
During training, we use datasets comprising 2D MR image slices (axial, coronal, and sagittal) along with their lesion masks~(either from manual annotation or automatic segmentation). 
Given a 2D MR image ${x_0}$ and its corresponding lesion mask ${M^\text{target}}$, a forward diffusion process $\{{x_{1:T}}\}$ is generated by gradually adding Gaussian noise to ${x_0}$ with a pre-defined mean and variance schedule~\cite{ho2020denoising}. 
During reverse diffusion, we use a U-Net-based denoiser to predict the noise between two adjacent time steps $t$ and $t-1$.
The DDPM takes both ${x_t}$ and ${M^\text{target}}$ as input, and predicts ${\hat{x}_{t-1}}$, which is a slightly denoised version of ${x}_t$.
The conditioning mask ${M^\text{target}}$ represents the desired lesion mask of the synthetic image, and it is used to guide the DDPM for either lesion filling~(${M^\text{target}} = 0$) or lesion generation. 
Once trained, the DDPM can generate lesion-free images by providing an all-zero $M^\text{target}$. 
It can also generate synthetic lesion images by providing the desired lesion mask.

\subsection{Inference: Lesion filling and generation}
To preserve the pixels outside the region of interest, we introduce another mask $M^{\text{repaint}}$. 
Only pixels within this mask will be updated for lesion filling or generation at inference time.
Inspired by the RePaint method~\cite{lugmayr2022repaint}, we introduce an additional step after the regular DDPM reverse diffusion step. 
Once we have the predicted denoised image ${\hat{x}_{t-1}}$ from time $t$, it is then updated by mixing ${\hat{x}_{t-1}}$ with the true noisy image $x_{t-1}$ calculated from the forward diffusion process.
The updated image, $\hat{x}_{t-1}'$, is defined as
\begin{align}
\label{eq:mix}
    {\hat{x}_{t-1}}' = {\hat{x}_{t-1}} \cdot {M^\text{repaint}} + {x_{t-1}} \cdot ({1} - {M^\text{repaint}}).
\end{align}
This guarantees that pixels outside $M^\text{repaint}$ are preserved, and only pixels within the mask are updated.
For better consistency between lesion and non-lesion regions, after updating ${\hat{x}_{t-1}}$ with ${\hat{x}_{t-1}}'$ using Eq.~\ref{eq:mix}, ${x_t}$ is regenerated from ${\hat{x}_{t-1}}$ with one-step forward diffusion and then run through the reverse process at timestep $t$ to `repaint' lesions~\cite{lugmayr2022repaint}.
This repaint process can be repeated several times, with the number of repetitions as a hyperparameter.

\noindent \textbf{Lesion filling:} 
Figure~\ref{fig:figure1} shows the overall diagram of inference. 
For lesion filling, the goal is to replace pixels of image $x_0$ within $M^\text{repaint}$ with healthy-looking pixels. 
Therefore, $M^{\text{target}}$ is set to be all-zero to guide the DDPM to generate lesion-free images.

\noindent \textbf{Lesion generation:} Similarly, lesion generation during inference involves replacing healthy-looking pixels with lesion-looking pixels.
This replacement is controlled by $M^\text{repaint}$. 
To guide the DDPM to generate lesion looking images, $M^\text{target}$ is set to $M^\text{repaint}$ to indicate the desired lesion masks.

\noindent \textbf{Inference acceleration with DDIM:}
To accelerate the inference process, we apply a DDIM with a sub-sequence $[\tau_1,..., \tau_S]$ of $[1,..., T]$ for reverse sampling ($S \ll T$).
Since the foward process is no longer Markovian in DDIM, we regenerate ${x_t}$ for lesion repaint using a inpainted ${\hat{x}_0}'$ instead of ${\hat{x}_{t-1}}'$ in Eq.~\ref{eq:mix}.
Inpainting ${\hat{x}_0}$ is similar to Eq.~\ref{eq:mix} with $t=0$: ${\hat{x}_0}' = {\hat{x}_0} \cdot {M^\text{repaint}} + {x_0} \cdot ({1} - {M^\text{repaint}})$, where $\hat{x}_0$ on the right-hand side is also estimated from the reverse process at timestep $t$.

\noindent \textbf{Additional forward and reverse sampling:}
To further improve consistency between non-lesion and inpainted regions, the estimated ${\hat{x}_0}$ after DDIM reverse sampling undergoes additional forward diffusion and reverse sampling (using the DDIM scheme) process:
$ {\hat{x}_0} 
    \xrightarrow{\text{Diffusion}} 
    {\hat{x}}_{\tau_I}  \xrightarrow{\text{Reverse}} 
    {\hat{x}_0}
$
with $\tau_I \in [\tau_1,...\tau_S]$ as a small timestep used to inject a small amount of Gaussian noise into ${\hat{x}_0}$ for additional reverse sampling.

\section{EXPERIMENTS}
\label{s:experiments}

\subsection{Setup}

\noindent \textbf{Dataset:}
The training dataset includes T1w and FLAIR images of 66 MS subjects from 8 private sites and their corresponding lesion masks, resulting in 26,800 and 26,000 2D slices of T1w and FLAIR images, respectively, in axial, coronal, or sagittal planes.
The test dataset for lesion filling includes T1w and FLAIR images along with the corresponding lesion masks of 15 MS subjects from a different site.
The test dataset for lesion synthesis includes T1w and FLAIR images of 10 healthy subjects and lesion masks of 10 MS subjects.
Each 3D lesion mask was intersected with the corresponding white matter mask of a healthy subject to generate plausible lesion masks for lesion synthesis.
All MS lesion masks were generated using SELF~\cite{zhang2024isbi}, which was trained on both the 2015 ISBI challenge training data (21 longitudinal multi-contrast images from 5 subjects)~\cite{carass2017ni} and in-house data with manual lesion delineation (20 multi-contrast images from 10 subjects and 2 sites).

\noindent\textbf{Training and Inference:}
The reverse process is modeled using a time-embedded U-Net backbone with $T=1000$ time steps. 
A cosine noise schedule of the forward process was applied for improved image quality~\cite{nichol2021improved}.
Training used the Adam optimizer with an initial learning rate of $3\times10^{-4}$, a batch size of 32 randomly selected from 2D slices in the three cardinal planes, and was conducted over 300 epochs.
DDIM inference used a subset of the original time steps, with every 10$^{\text{th}}$ step included.
Lesion repainting was repeated twice at each timestep.

\subsection{Results}
\begin{figure}[!t]
	\centering
    \includegraphics[width=0.75\columnwidth]{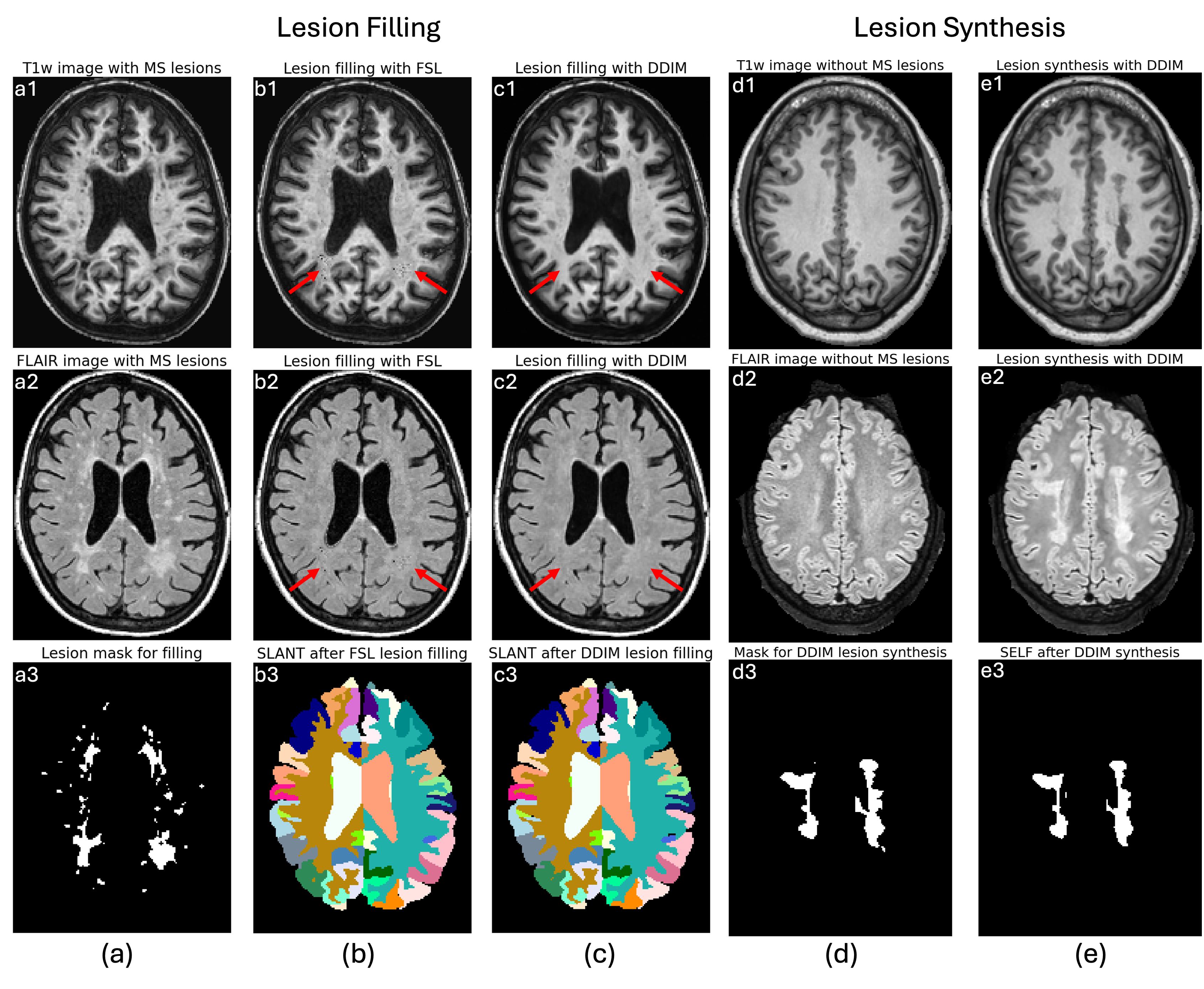}
	\caption{\textbf{(a)-(c): Lesion filling results.} \textbf{(a1-3)}:~Original T1w, FLAIR, and lesion mask. \textbf{(b1-3)}:~Lesion filling of T1w and FLAIR by FSL and the corresponding SLANT brain parcellation on lesion-filled T1w image. \textbf{(c1-3)}:~Lesion filling results using the proposed method and the corresponding brain parcellation.
    \textbf{(d)-(e):~Lesion Synthesis results.}~\textbf{(d1-3)}:~T1w and FLAIR images from a healthy subject, and the binary lesion mask for the lesion synthesis task. \textbf{(e1-3)}:~Synthetic T1w and FLAIR images with lesions, and the corresponding lesion segmentation mask.
	}
	\label{fig:figure2}
\end{figure}

\noindent \textbf{Lesion Filling:}
In Fig.~\ref{fig:figure2},~(a) shows T1w, FLAIR, and lesion mask images of a representative MS subject.
(b) shows FSL lesion filling~\cite{battaglini2012evaluating} results on the T1w, FLAIR, and the SLANT~\cite{huo20193d} brain parcellation map derived from the corresponding lesion-filled T1w image, while~(c) shows results for our proposed DDPM lesion filling method.
The red arrows, in Fig.~\ref{fig:figure2}, highlight the regions of lesion infilling.
Mean Dice similarity scores~(DSCs) with standard deviations for regions of interest~(ROIs) between SLANT using FSL and the proposed filling methods across 15 test subjects were as follows: Ventricles: 0.932 $\pm$ 0.144, Cerebral White Matter: 0.915 $\pm$ 0.174, Brainstem: 0.971 $\pm$ 0.053, Caudate: 0.864 $\pm$ 0.301, Cerebellar Gray Matter: 0.949 $\pm$ 0.102, Cerebellar White Matter: 0.901 $\pm$ 0.213, Putamen: 0.854 $\pm$ 0.334, Thalamus: 0.936 $\pm$ 0.161, and Cerebrum Gray Matter: 0.908 $\pm$ 0.188. 

\noindent\textbf{Lesion Synthesis:}
In Fig.~\ref{fig:figure2}, (d) shows the T1w and FLAIR images of a representative healthy subject, along with the corresponding binary mask used for lesion synthesis.
(e) shows the T1w and FLAIR images with synthetic lesions, and the corresponding SELF lesion segmentation mask after applying SELF on the synthetic T1w and FLAIR pair.
Realistic lesions were generated with white matter hypo-intensity on the T1w image and hyper-intensity on the FLAIR image, see Figs.~\ref{fig:figure2}(e1) and~(e2) respectively.
Lesion segmentation (Fig.~\ref{fig:figure2}(e3)) from the synthetic T1w and FLAIR images shows a generally consistent lesion mask, compared to the original lesion mask used for synthesis, see Fig.~\ref{fig:figure2}(d3).
A mean DSC of 0.832~($\pm$ 0.049) across 10 test subjects was achieved between the segmented lesion mask from synthetic images and the original lesion mask used for synthesis.

\section{DISCUSSION AND CONCLUSION}
\label{s:discussion_and_conclusion}
In this work, we present preliminary results of bi-directional lesion filling and synthesis using a single trained DDPM model with DDIM-based inference acceleration.
In the lesion filling task, DDIM generates regions filled with normal appearing white matter tissue contrast (red arrows in Fig.~\ref{fig:figure2}(c)), which can be attributed to the DDPM training process using reliable lesion masks that correspond to lesion regions only.
As a result, after DDPM training, the non-lesion regions will be filled with normal appearing white matter resembling that of the training dataset.
For the downstream whole-brain parcellation, some inconsistencies between the SLANT results of FSL lesion filling and our lesion filling were observed, this is an area for further analysis.
In the lesion synthesis task, DDIM generates realistic lesions in both T1w and FLAIR images (Figs.~\ref{fig:figure2}(e1) and~(e2)) with high DSCs after lesion segmentation on the synthetic images.
Such synthetic images with ground truth lesion masks could be useful for training more domain-generalizable lesion segmentation models~\cite{zhang2024spie}.

\section{PLANNED WORK}
\label{s:discussion_and_conclusion}
The full paper will include further validation of lesion filling accuracy and additional exploration of using synthetic multi-contrast MR images with lesions to augment training datasets for lesion segmentation models.

\centerline{\textbf{This work has not been submitted for publication or presentation elsewhere.}}

\section*{Acknowledgments}
This work was partially supported by the National Science Foundation Graduate Research Fellowship under Grant No. DGE-1746891 (Remedios).
Development is partially supported by CDMRP W81XWH2010912~(Prince), NIH R01 CA253923~(Landman), NIH R01 CA275015~(Landman), the NMSS grant RG-1507-05243~(Pham) and PCORI grant MS-1610-37115~(Newsome and Mowry).
The statements in this publication are solely the responsibility of the authors and do not necessarily represent the views of PCORI, its Board of Governors or Methodology Committee.

\bibliography{report} 
\bibliographystyle{spiebib} 

\end{document}